\documentclass[12pt, prd, showpacs]{revtex4}
\usepackage{amssymb}
\usepackage{amsmath}

\input{tcilatex}

\begin{document}

\title{General properties of the Penrose process with neutral particles in
the equatorial plane}
\author{O. B. Zaslavskii}
\affiliation{Department of Physics and Technology, Kharkov V.N. Karazin National
University, 4 Svoboda Square, Kharkov 61022, Ukraine}
\email{zaslav@ukr.net }

\begin{abstract}
We consider the background of a rotating axially symmetric black hole. Let
particle 0 decay to two fragments 1 and 2 in the direction parallel to that
of particle 0. It is shown that if decay occurs inside the ergoregion, both
particles 1 and 2 move in the same direction as particle 0. For the
scenario, when decay happens in the turning point of all three particles, we
find the condition when angular momenta of both particles 1 and 2 have the
same sign. We elucidate the relation between the approach of Wald that
imposes constraint on maximum and minimum energies of fragments and our
approach. In doing so, we express the results in terms of characteristics of
particle 0 and all particle masses. The conditions of the maximum efficiency
depending on the relation between masses is discussed. We find an explicit
expression for angular momenta of particles 1 and 2. We discuss also
particle decay for static black holes, when the Penrose process is
impossible. Because of the absence of the ergoregion in the static case,
scenarios of decay for static black holes can significantly differ from
those in the rotating background.
\end{abstract}

\maketitle

\section{Introduction}

The Penrose process (PP) was discovered more than 50 years ago \cite{pen}, 
\cite{pen2} but currently it is experiencing a new wave of interest. It
concerns both theoretical and astrophysically relevant issues. It is an
essential ingredient in the discussion of energy extraction from black holes
in high energy scenarios, in particular in a magnetic field \cite{dad1} - 
\cite{rufkerr}. Originally, the PP was found for rotating black holes.
Meanwhile, it was shown later that a similar phenomenon exists also for
static electrically charged black holes \cite{ruf}, \cite{den}. Energy
extraction for combination of rotation and electric charge is considered in 
\cite{fh}. Moreover, some analogs of PP exist even in the flat space-times 
\cite{df}, \cite{flat}, \cite{naked}. New types of PP were found and
elaborated in a recent decade. This includes the collisional PP \cite{shn}
playing a crucial role in high energy collisions between particles near
black holes, wormholes and singularities. It is an important ingredient for
the Ba\~{n}ados-Silk-West (BSW) effects and its analogues \cite{ban} - \cite%
{pir3}. Another example is a confined Penrose process that takes place for
massive particles inside a cavity or reflecting shells \cite{conf}, \cite%
{myconf}, being a counterpart of wave amplification and superradiance \cite%
{st1}- \cite{pani}. A concise review with a useful list of references can be
found in the Introduction of \cite{win}.

All this makes it quite necessary to have well-elaborated and
model-independent formulas describing the PP. Meanwhile, strange as it might
seem, in spite of an investigation of the PP in a wide context including
even quite sophisticated models \cite{j}, there are some gaps in basics of
the PP. In a particular, a quite complicated method of investigation of the
PP became rather popular (see Sec. \ref{two}), although more simple and
straightforward one \cite{wald} went into the shadows.

The goal of the present paper is twofold. (i) We elaborate basic formalism
describing PP for a rotating axially symmetric quite generic stationary
background. (ii) We give classification of possible scenarios and describe
some generic features of processes in the ergosphere that remained unnoticed
up to now. In particular, we trace carefully which scenario is possible (or
impossible) if decay happens near the horizon. To this end, we list in
detail different scenarios, indicating explicitly where the "plus" or
"minus" sign should stand. As by product, we dispel some incorrect
statements made in the literature during past several years. Also, we stress
that the properties of decay in the static background is not just a
particular case of a more general case in rotating one. This comparison
reveals some qualitative difference, especially for decay near the horizon,
where one should be careful.

The paper is organized as follows. In Sec. \ref{eqs} we list the general
form of a metric and equations of particle motion. In Sec. \ref{gen} we give
general relations between characteristics of a parent particle 0 and
particles 1 and 2 to which it decays. In Sec. \ref{class} we suggest
classification of possible scenarios of decay. In Sec. \ref{along} we
discuss general features of a scenario in which particles 1 and 2 are
ejected parallel to the direction of motion of particle 0. In Sec. \ref{prop}
we discuss properties of corresponding scenarios inside and outside the
ergoregion. In Sec. \ref{maxim} we show that the equations governing motion
of particles in the scenarios of parallel ejection can be also obtained from
the requirement of maximization (minimization) of the energy of ejected
particles. In Sec. \ref{wald} we demonstrate equivalence between the derived
formulas and Wald bounds but instead of velocity in the center of mass frame
we use the values of particle masses, thus reformulating the aforementioned
bounds in other terms. In Sec. \ref{tp} we consider a special kind of
scenario when decay occurs in the turning point for all three particles. For
the aforementioned scenario, in Sec. \ref{signs} we consider possible signs
of angular momenta of particles 1\ and 2 in the ergoregion. In Sec. \ref{thr}
we formulate the threshold for the Penrose process in terms of particle
masses. In Sec. \ref{efrot} we discuss general features of the efficiency of
the decay process. In Sec. \ref{stat} we discuss peculiarities of the decay
for static black holes. In Sec. \ref{two} we compare our approach and
another one rather popular in literature. In Sec. \ref{sum} we summarize the
results obtained.

\section{Equations of motion\label{eqs}}

Let us consider the metric%
\begin{equation}
ds^{2}=-N^{2}dt^{2}+g_{\phi }(d\phi -\omega dt)^{2}+\frac{dr^{2}}{A}%
+g_{\theta }d\theta ^{2},  \label{met}
\end{equation}%
where we used notations $g_{\phi }\equiv g_{\phi \phi }$ and $g_{\theta
}\equiv g_{\theta \theta }$ for brevity. We restrict ourselves by stationary
and axially symmetric metrics only, so we assume that the metric
coefficients do not depend on $\phi $ and $t$.

We assume the symmetry with respect to the equatorial plane $\theta =\frac{%
\pi }{2}$, so that the metric coefficient depend algebraically on $\sin
^{2}\theta $. Then, there exist geodesics just in this plane, and we
restrict ourselves by particle motion within this plane. Then, equations of
motion for a free particle read%
\begin{equation}
m\dot{t}=\frac{X}{N^{2}}\text{,}  \label{t}
\end{equation}%
\begin{equation}
p^{r}\equiv m\dot{r}=\sigma \frac{\sqrt{A}}{N}P\text{, }P=\sqrt{X^{2}-\tilde{%
m}^{2}N^{2}},  \label{P}
\end{equation}%
\begin{equation}
X=E-\omega L\text{,}  \label{x}
\end{equation}%
\begin{equation}
\tilde{m}^{2}=m^{2}+\frac{L^{2}}{g_{\phi }},  \label{m}
\end{equation}%
\begin{equation}
m\dot{\phi}=\frac{L}{g_{\phi }}+\frac{\omega X}{N^{2}}=\frac{\omega E}{N^{2}}%
-\frac{Lg_{00}}{N^{2}g_{\phi }},  \label{phi}
\end{equation}%
where dot denotes differentiation with respect to the proper time. The
direction of motion is characterized by a quantity $\sigma $, where $\sigma
=+1$ for motion in the outward direction and $\sigma =-1$ for the inward
case. Here, $E$ is the particle energy, $L$ being angular \ momentum, $m$
its mass. Eqs. (\ref{t}) and (\ref{phi}) are direction consequences of the
conservation of the energy and angular momentum, eq. (\ref{P}) uses the
normalization conditions for the four-velocity $u_{\mu }u^{\mu }=-1$.

As a consequence of the above equations, the angular velocity%
\begin{equation}
\Omega \equiv \frac{d\phi }{dt}=\omega +\frac{LN^{2}}{g_{\phi }X}.
\label{ome}
\end{equation}%
The requirement $\dot{t}>0$ leads to the forward-in-time condition%
\begin{equation}
X\geq 0.  \label{ft}
\end{equation}

For the four-velocity one can write in coordinates $(t,r,\phi )$%
\begin{equation}
u^{\mu }=(\frac{X}{mN^{2}}\text{, }\sigma \frac{\sqrt{A}}{mN}P,\frac{L}{%
mg_{\phi }}+\frac{\omega X}{mN^{2}})\text{.}  \label{u}
\end{equation}

We omit the component $u^{\theta }=0$.\ \ \ \ \ \ \ \ \ \ \ \ \ \ \ \ \ \ \
\ \ \ \ \ \ \ \ \ \ \ \ \ \ \ \ \ \ \ \ \ \ \ \ \ \ \ \ \ \ \ \ \ \ \ \ \ \
\ \ \ \ \ \ \ \ \ \ \ \ \ \ \ \ \ \ \ \ \ \ \ \ \ \ \ \ \ \ \ \ \ \ \ \ \ \
\ \ \ \ \ \ \ \ \ \ \ \ \ \ \ \ \ \ \ \ \ \ \ \ \ \ \ \ \ \ \ \ \ \ \ \ \ \
\ \ \ \ \ \ \ \ \ \ \ \ \ \ \ \ \ \ \ \ \ \ \ \ \ \ \ \ \ \ \ \ \ \ \ \ \ \
\ \ \ \ \ \ \ \ \ \ \ \ \ \ \ \ \ \ \ \ \ \ \ \ \ \ \ \ \ \ \ \ \ \ \ \ \ \
\ \ \ \ \ \ \ \ \ \ \ \ \ \ \ \ \ \ \ \ \ \ \ \ \ \ \ \ \ \ \ \ \ \ \ \ \ \
\ \ \ \ \ \ \ \ \ \ \ \ \ \ \ \ \ \ \ \ \ \ \ \ \ \ \ \ \ \ \ \ \ \ \ \ \ \
\ \ \ \ \ \ \ \ \ \ \ \ \ \ \ \ \ \ \ \ \ \ \ \ \ \ \ \ \ \ \ \ \ \ \ \ \ \
\ \ \ \ \ \ \ \ \ \ \ \ \ \ \ \ \ \ \ \ \ \ \ \ \ \ \ \ \ \ \ \ \ \ \ \ \ \
\ \ \ \ \ \ \ \ \ \ \ \ \ \ \ \ \ \ \ \ \ \ \ \ \ \ \ \ \ \ \ \ \ \ \ \ \ \
\ \ \ \ \ \ \ \ \ \ \ \ \ \ \ \ \ \ \ \ \ \ \ \ \ \ \ \ \ \ \ \ \ \ \ \ \ \
\ \ \ \ \ \ \ \ \ \ \ \ \ \ \ \ \ \ \ \ \ \ \ \ \ \ \ \ \ \ \ \ \ \ \ \ \ \
\ \ \ \ \ \ \ \ \ \ \ \ \ \ \ \ \ \ \ \ \ \ \ \ \ \ \ \ \ \ \ \ \ \ \ \ \ \
\ \ \ \ \ \ \ \ \ \ \ \ \ \ \ \ \ \ \ \ \ \ \ \ \ \ \ \ \ \ \ \ \ \ \ \ \ \
\ \ \ \ \ \ \ \ \ \ \ \ \ \ \ \ \ \ \ \ \ \ \ \ \ \ \ \ \ \ \ \ \ \ \ \ \ \
\ \ \ \ \ \ \ \ \ \ \ \ \ \ \ \ \ \ \ \ \ \ \ \ \ \ \ \ \ \ \ \ \ \ \ \ \ \
\ \ \ \ \ \ \ \ \ \ \ \ \ \ \ \ \ \ \ \ \ \ \ \ \ \ \ \ \ \ \ \ \ \ \ \ \ \
\ \ \ \ \ \ \ \ \ \ \ \ \ \ \ \ \ \ \ \ \ \ \ \ \ \ \ \ \ \ \ \ \ \ \ \ \ \
\ \ \ \ \ \ \ \ \ \ \ \ \ \ \ \ \ \ \ \ \ \ \ \ \ \ \ \ \ \ \ \ \ \ \ \ \ \
\ \ \ \ \ \ \ \ \ \ \ \ \ \ \ \ \ \ \ \ \ \ \ \ \ \ \ \ \ \ \ \ \ \ \ \ \ \
\ \ \ \ \ \ \ \ \ \ \ \ \ \ \ \ \ \ \ \ \ \ \ \ \ \ \ \ \ \ \ \ \ \ \ \ \ \
\ \ \ \ \ \ \ \ \ \ \ \ \ \ \ \ \ \ \ \ \ \ \ \ \ \ \ \ \ \ \ \ \ \ \ \ \ \
\ \ \ \ \ \ \ \ \ \ \ \ \ \ \ \ \ \ \ \ \ \ \ \ \ \ \ \ \ \ \ \ \ \ \ \ \ \
\ \ \ \ \ \ \ \ \ \ \ \ \ \ \ \ \ \ \ \ \ \ \ \ \ \ \ \ \ \ \ \ \ \ \ \ \ \
\ \ \ \ \ \ \ \ \ \ \ \ \ \ \ \ \ \ \ \ \ \ \ \ \ \ \ \ \ \ \ \ \ \ \ \ \ \
\ \ \ \ \ \ \ \ \ \ \ \ \ \ \ \ \ \ \ \ \ \ \ \ \ \ \ \ \ \ \ \ \ \ \ \ \ \
\ \ \ \ \ \ \ \ \ \ \ \ \ \ \ \ \ \ \ \ \ \ \ \ \ \ \ \ \ \ \ \ \ \ \ \ \ \
\ \ \ \ \ \ \ \ \ \ \ \ \ \ \ \ \ \ \ \ \ \ \ \ \ \ \ \ \ \ \ \ \ \ \ \ \ \
\ \ \ \ \ \ \ \ \ \ \ \ \ \ \ \ \ \ \ \ \ \ \ \ \ \ \ \ \ \ \ \ \ \ \ \ \ \
\ \ \ \ \ \ \ \ \ \ \ \ \ \ \ \ \ \ \ \ \ \ \ \ \ \ \ \ \ \ \ \ \ \ \ \ \ \
\ \ \ \ \ \ \ \ \ \ \ \ \ \ \ \ \ \ \ \ \ \ \ \ \ \ \ \ \ \ \ \ \ \ \ \ \ \
\ \ \ \ \ \ \ \ \ \ \ \ \ \ \ \ \ \ \ \ \ \ \ \ \ \ \ \ \ \ \ \ \ \ \ \ \ \
\ \ \ \ \ \ \ \ \ \ \ \ \ \ \ \ \ \ \ \ \ \ \ \ \ \ \ \ \ \ \ \ \ \ \ \ \ \
\ \ \ \ \ \ \ \ \ \ \ \ \ \ \ \ \ \ \ \ \ \ \ \ \ \ \ \ \ \ \ \ \ \ \ \ \ \
\ \ \ \ \ \ \ \ \ \ \ \ \ \ \ \ \ \ \ \ \ \ \ \ \ \ \ \ \ \ \ \ \ \ \ \ \ \
\ \ \ \ \ \ \ \ \ \ \ \ \ \ \ \ \ \ \ \ \ \ \ \ \ \ \ \ \ \ \ \ \ \ \ \ \ \
\ \ \ \ \ \ \ \ \ \ \ \ \ \ \ \ \ \ \ \ \ \ \ \ \ \ \ \ \ \ \ \ \ \ \ \ \ \
\ \ \ \ \ \ \ \ \ \ \ \ \ \ \ \ \ \ \ \ \ \ \ \ \ \ \ \ \ \ \ \ \ \ \ \ \ \
\ \ \ \ \ \ \ \ \ \ \ \ \ \ \ \ \ \ \ \ \ \ \ \ \ \ \ \ \ \ \ \ \ \ \ \ \ \
\ \ \ \ \ \ \ \ \ \ \ \ \ \ \ \ \ \ \ \ \ \ \ \ \ \ \ \ \ \ \ \ \ \ \ \ \ \
\ \ \ \ \ \ \ \ \ \ \ \ \ \ \ \ \ \ \ \ \ \ \ \ \ \ \ \ \ \ \ \ \ \ \ \ \ \
\ \ \ \ \ \ \ \ \ \ \ \ \ \ \ \ \ \ \ \ \ \ \ \ \ \ \ \ \ \ \ \ \ \ \ \ \ \
\ \ \ \ \ \ \ \ \ \ \ \ \ \ \ \ \ \ \ \ \ \ \ \ \ \ \ \ \ \ \ \ \ \ \ \ \ \
\ \ \ \ \ \ \ \ \ \ \ \ \ \ \ \ \ \ \ \ \ \ \ \ \ \ \ \ \ \ \ \ \ \ \ \ \ \
\ \ \ \ \ \ \ \ \ \ \ \ \ \ \ \ \ \ \ \ \ \ \ \ \ \ \ \ \ \ \ \ \ \ \ \ \ \
\ \ \ \ \ \ \ \ \ \ \ \ \ \ \ \ \ \ \ \ \ \ \ \ \ \ \ \ \ \ \ \ \ \ \ \ \ \
\ \ \ \ \ \ \ \ \ \ \ \ \ \ \ \ \ \ \ \ \ \ \ \ \ \ \ \ \ \ \ \ \ \ \ \ \ \
\ \ \ \ \ \ \ \ \ \ \ \ \ \ \ \ \ \ \ \ \ \ \ \ \ \ \ \ \ \ \ \ \ \ \ \ \ \
\ \ \ \ \ \ \ \ \ \ \ \ \ \ \ \ \ \ \ \ \ \ \ \ \ \ \ \ \ \ \ \ \ \ \ \ \ \
\ \ \ \ \ \ \ \ \ \ \ \ \ \ \ \ \ \ \ \ \ \ \ \ \ \ \ \ \ \ \ \ \ \ \ \ \ \
\ \ \ \ \ \ \ \ \ \ \ \ \ \ \ \ \ \ \ \ \ \ \ \ \ \ \ \ \ \ \ \ \ \ \ \ \ \
\ \ \ \ \ \ \ \ \ \ \ \ \ \ \ \ \ \ \ \ \ \ \ \ \ \ \ \ \ \ \ \ \ \ \ \ \ \
\ \ \ \ \ \ \ \ \ \ \ \ \ \ \ \ \ \ \ \ \ \ \ \ \ \ \ \ \ \ \ \ \ \ \ \ \ \
\ \ \ \ \ \ \ \ \ \ \ \ \ \ \ \ \ \ \ \ \ \ \ \ \ \ \ \ \ \ \ \ \ \ \ \ \ \
\ \ \ \ \ \ \ \ \ \ \ \ \ \ \ \ \ \ \ \ \ \ \ \ \ \ \ \ \ \ \ \ \ \ \ \ \ \
\ \ \ \ \ \ \ \ \ \ \ \ \ \ \ \ \ \ \ \ \ \ \ \ \ \ \ \ \ \ \ \ \ \ \ \ \ \
\ \ \ \ \ \ \ \ \ \ \ \ \ \ \ \ \ \ \ \ \ \ \ \ \ \ \ \ \ \ \ \ \ \ \ \ \ \
\ \ \ \ \ \ \ \ \ \ \ \ \ \ \ \ \ \ \ \ \ \ \ \ \ \ \ \ \ \ \ \ \ \ \ \ \ \
\ \ \ \ \ \ \ \ \ \ \ \ \ \ \ \ \ \ \ \ \ \ \ \ \ \ \ \ \ \ \ \ \ \ \ \ \ \
\ \ \ \ \ \ \ \ \ \ \ \ \ \ \ \ \ \ \ \ \ \ \ \ \ \ \ \ \ \ \ \ \ \ \ \ \ \
\ \ \ \ \ \ \ \ \ \ \ \ \ \ \ \ \ \ \ \ \ \ \ \ \ \ \ \ \ \ \ \ \ \ \ \ \ \
\ \ \ \ \ \ \ \ \ \ \ \ \ \ \ \ \ \ \ \ \ \ \ \ \ \ \ \ \ \ \ \ \ \ \ \ \ \
\ \ \ \ \ \ \ \ \ \ \ \ \ \ \ \ \ \ \ \ \ \ \ \ \ \ \ \ \ \ \ \ \ \ \ \ \ \
\ \ \ \ \ \ \ \ \ \ \ \ \ \ \ \ \ \ \ \ \ \ \ \ \ \ \ \ \ \ \ \ \ \ \ \ \ \
\ \ \ \ \ \ \ \ \ \ \ \ \ \ \ \ \ \ \ \ \ \ \ \ \ \ \ \ \ \ \ \ \ \ \ \ \ \
\ \ \ \ \ \ \ \ \ \ \ \ \ \ \ \ \ \ \ \ \ \ \ \ \ \ \ \ \ \ \ \ \ \ \ \ \ \
\ \ \ \ \ \ \ \ \ \ \ \ \ \ \ \ \ \ \ \ \ \ \ \ \ \ \ \ \ \ \ \ \ \ \ \ \ \
\ \ \ \ \ \ \ \ \ \ \ \ \ \ \ \ \ \ \ \ \ \ \ \ \ \ \ \ \ \ \ \ \ \ \ \ \ \
\ \ \ \ \ \ \ \ \ \ \ \ \ \ \ \ \ \ \ \ \ \ \ \ \ \ \ \ \ \ \ \ \ \ \ \ \ \
\ \ \ \ \ \ \ \ \ \ \ \ \ \ \ \ \ \ \ \ \ \ \ \ \ \ \ \ \ \ \ \ \ \ \ \ \ \
\ \ \ \ \ \ \ \ \ \ \ \ \ \ \ \ \ \ \ \ \ \ \ \ \ \ \ \ \ \ \ \ \ \ \ \ \ \
\ \ \ 

\section{General setup \label{gen}}

Let in the point $r=r_{0}$ a parent particle 0 decay to particles 1 and 2.
Hereafter, we call it "scenario AP" (since this happens in an arbitrary
point). We assume the conservation laws in this point:%
\begin{equation}
E_{0}=E_{1}+E_{2}\text{,}  \label{e12}
\end{equation}%
\begin{equation}
L_{0}=L_{1}+L_{2}\text{,}  \label{L0}
\end{equation}%
\begin{equation}
p_{0}^{r}=p_{1}^{r}+p_{2}^{r}\text{,}  \label{pr}
\end{equation}%
whence%
\begin{equation}
X_{0}=X_{1}+X_{2}\text{.}  \label{x12}
\end{equation}

Then, using these equations and (\ref{P}), (\ref{x}), one can obtain%
\begin{equation}
X_{2}=\frac{1}{2\tilde{m}_{0}^{2}}\left( X_{0}\tilde{b}_{2}+P_{0}\delta 
\sqrt{\tilde{d}}\right) ,  \label{e2}
\end{equation}

\begin{equation}
X_{1}=\frac{1}{2\tilde{m}_{0}^{2}}\left( X_{0}\tilde{b}_{1}-P_{0}\delta 
\sqrt{\tilde{d}}\right) ,  \label{e1}
\end{equation}%
\begin{equation}
\tilde{b}_{1}=\tilde{m}_{0}^{2}+\tilde{m}_{1}^{2}-\tilde{m}_{2,}^{2},
\label{del}
\end{equation}%
\begin{equation}
\tilde{b}_{2}=\tilde{m}_{0}^{2}+\tilde{m}_{2}^{2}-\tilde{m}_{1,}^{2},
\label{b2t}
\end{equation}%
where $i=0,1,2$, $\delta =\pm 1$,%
\begin{equation}
\tilde{m}_{i}^{2}=m_{i}^{2}+\frac{L_{i}^{2}}{g_{\phi }}\text{,}  \label{mt}
\end{equation}%
\begin{equation}
\tilde{m}_{0}\geq \tilde{m}_{1}+\tilde{m}_{2}\text{.}  \label{m12t}
\end{equation}%
\begin{equation}
\tilde{d}=\tilde{b}_{1}^{2}-4\tilde{m}_{0}^{2}\tilde{m}_{1}^{2}=\tilde{b}%
_{2}^{2}-4\tilde{m}_{0}^{2}\tilde{m}_{2}^{2},  \label{D}
\end{equation}%
\begin{equation}
P_{2}=\left\vert \frac{P_{0}\tilde{b}_{2}+\delta X_{0}\sqrt{\tilde{d}}}{2%
\tilde{m}_{0}^{2}}\right\vert ,  \label{p2}
\end{equation}

\begin{equation}
P_{1}=\left\vert \frac{P_{0}\tilde{b}_{1}-\delta X_{0}\sqrt{\tilde{d}}}{2%
\tilde{m}_{0}^{2}}\right\vert \text{.}  \label{p1}
\end{equation}

Here, parameter $\delta $ appears formally in the course of solving
quadratic equations after eliminating square roots in (\ref{x12}). Its
presence shows that, for given characteristics of particle 0, there are two
different (in general, nonequivalent) scenarios with fixed masses of
particles 1 and 2.

A useful equality follows from (\ref{del}), (\ref{b2t}):%
\begin{equation}
\tilde{b}_{1}+\tilde{b}_{2}=2\tilde{m}_{0}^{2}.  \label{b12}
\end{equation}

The above expressions are similar to the results already obtained in
eqs.~(19) --- (30) of~\cite{centr} \ for particle collisions. The difference
is that now we are dealing with particle decay but the equations retain the
same structure. We use now particle labels 1 and 2 instead of 3 and 4
respectively in~\cite{centr}.

In what follows, we will exploit useful equalities

\begin{equation}
\tilde{b}_{1,2}=b_{1,2}+\frac{2L_{0}L_{1,2}}{g_{\phi }},  \label{btil}
\end{equation}%
\begin{equation}
\tilde{d}=d+4b_{1}\frac{L_{0}L_{1}}{g_{\phi }}-\frac{4L_{1}^{2}}{g_{\phi }}%
m_{0}^{2}-4\frac{L_{0}^{2}}{g_{\phi }}m_{1}^{2}.  \label{dtil}
\end{equation}

Equivalently,%
\begin{equation}
\tilde{d}=d+4b_{2}\frac{L_{0}L_{2}}{g_{\phi }}-\frac{4L_{2}^{2}}{g_{\phi }}%
m_{0}^{2}-4\frac{L_{0}^{2}}{g_{\phi }}m_{2}^{2}.
\end{equation}

Here,%
\begin{equation}
b_{1}=m_{0}^{2}+m_{1}^{2}-m_{2,}^{2},  \label{b}
\end{equation}%
\begin{equation}
b_{2}=m_{0}^{2}+m_{2}^{2}-m_{1,}^{2},  \label{b2}
\end{equation}%
\begin{equation}
d=b_{1}^{2}-4m_{0}^{2}m_{1}^{2}=b_{2}^{2}-4m_{0}^{2}m_{2}^{2}.  \label{d}
\end{equation}

\section{Classification of scenarios \label{class}}

We are mainly interested in the situation, when particle~0 moves from larger
radii to smaller ones, so $\sigma _{0}=-1$. Let, say, the metric describe a
black hole. Then, it follows from (\ref{pr}) that at least one of daughter
particles moves towards a black hole. By definition, we assume that $\sigma
_{1}=-1$. For particle 2, we cannot fix $\sigma _{2}$ beforehand. The
conservation law gives us%
\begin{equation}
-P_{0}=-P_{1}+\sigma _{2}P_{2}\text{.}  \label{012}
\end{equation}%
Using general expressions (\ref{p2}) and (\ref{p1}) as well as (\ref{012}),
we can enumerate all possible scenarios.

It is convenient to describe a scenario as a four-component vector $(\sigma
_{2}$,$h_{2}$,$h_{1},\delta )$ in the parametric space. Here,

\begin{equation}
h_{1,2}=sign(\tilde{b}_{1,2}N-2\tilde{m}_{1,2}X_{0}),  \label{h12}
\end{equation}%
this quantity arises when different signs inside the absolute values in (\ref%
{p2}), (\ref{p1}) are taken into account.

Then, from the conservation laws we have the following types of scenarios:\ 
\begin{equation}
\text{I }(+,+,h_{1},-)\text{, II }(-,h_{2},-,+),\text{ III }(-,-,h_{1},-)%
\text{. }
\end{equation}

They are equivalent to scenarios 
\begin{equation}
\text{1 }(+,+,+,-),\text{ 2 }(+,+,-,-),\text{3 }(-,-,-,+)\text{, 4 }(-,+,-,+)%
\text{, 5 }(-,-,-,-)\text{, 6 }(-,-,+,-).
\end{equation}%
from \cite{centr}. (But a reader should bear in mind that there is a typo in
scenario 6 in \cite{centr}).

More explicitly, if $\delta =-1,$%
\begin{equation}
P_{2}=\frac{(X_{0}\sqrt{\tilde{d}}-P_{0}\tilde{b}_{2})h_{2}}{2\tilde{m}%
_{0}^{2}}\text{,}  \label{p2t}
\end{equation}%
\begin{equation}
P_{1}=\frac{P_{0}\tilde{b}_{1}+X_{0}\sqrt{\tilde{d}}}{2\tilde{m}_{0}^{2}}.
\label{p1t}
\end{equation}%
If $\delta =+1,$%
\begin{equation}
P_{2}=\frac{X_{0}\sqrt{\tilde{d}}+P_{0}\tilde{b}_{2}}{2\tilde{m}_{0}^{2}}%
\text{,}  \label{P2t}
\end{equation}%
\begin{equation}
P_{1}=\frac{X_{0}\sqrt{\tilde{d}}-P_{0}\tilde{b}_{1}}{2\tilde{m}_{0}^{2}}%
h_{1}.  \label{P1t}
\end{equation}

If $\sigma =-1$, $P_{1}+P_{2}=P_{0}$ and we must take $\delta =-1$, $h_{2}$ $%
=-1$ or $\delta =+1$, $h_{1}=-1$

If $\sigma =+1$, $P_{1}-P_{2}=P_{0}$ and we must take $\delta =-1$, $%
h_{2}=+1 $

\section{Ejection along the direction of motion\label{along}}

The above equations are valid in a general case. They allow us to find $%
E_{1},E_{2}$ and $P_{1},P_{2}$ in terms of quantities characterizing initial
particle 0, plus one more free parameter (say, $L_{2}\,$). This uncertainty
disappears, if we impose an additional constraint. Let us assume that in the
coordinate frame (\ref{met}) particles 1 and 2 are ejected parallel to the
direction of motion of particle 0. This means that the tangent vector to the
trajectory keeps the same direction after decay. Therefore, the ratio $%
u^{r}/u^{\phi }$ remains the same. Correspondingly, for particle 1 that
moves in the same direction as particle 0 ($\sigma _{1}=\sigma _{0}=-1$) we
have from (\ref{u}) 
\begin{equation}
\frac{P_{1}}{P_{0}}=\frac{\frac{L_{1}}{g_{\phi }}+\frac{\omega X_{1}}{N^{2}}%
}{\frac{L_{0}}{g_{\phi }}+\frac{\omega X_{0}}{N^{2}}},  \label{tr}
\end{equation}%
where subscript denotes a particle's label. For a static metric, $\omega =0$%
, whence%
\begin{equation}
\frac{P_{1}}{P_{0}}=\frac{L_{1}}{L_{0}}\text{,}
\end{equation}%
and we return to the situation considered in \cite{rocket} for the
Schwarzschild metric. However, for $\omega \neq 0,$ a more general relation (%
\ref{tr}) should hold. It follows from it that%
\begin{equation}
L_{1}=\frac{P_{1}}{P_{0}}(L_{0}+g_{\phi }\frac{\omega }{N^{2}})-\frac{\omega 
}{N^{2}}g_{\phi }X_{1}\text{.}  \label{eq}
\end{equation}%
Let us consider, say, scenario I. It follows from (\ref{p1}), (\ref{e1}) that%
\begin{equation}
L_{1}=\frac{L_{0}}{2\tilde{m}_{0}^{2}}\tilde{b}_{1}+\frac{\sqrt{\tilde{d}}}{2%
\tilde{m}_{0}^{2}P_{0}}(L_{0}X_{0}+g_{\phi }\omega \tilde{m}_{0}^{2})\text{,}
\label{L1+}
\end{equation}%
\begin{equation}
L_{2}=\frac{L_{0}}{2\tilde{m}_{0}^{2}}\tilde{b}_{2}-\frac{\sqrt{\tilde{d}}}{2%
\tilde{m}_{0}^{2}P_{0}}(L_{0}X_{0}+g_{\phi }\omega \tilde{m}_{0}^{2})\text{.}
\label{L2+}
\end{equation}

This is not the end of story since the quantities $\tilde{b}_{1,2}$ and $%
\tilde{d}$ themselves contain $L_{1,2}$.

Our goal is, given $E_{0},L_{0}$, $m_{0},m_{1},m_{2,}$ to find $E_{1}$, $%
E_{2}$, $L_{1}$, $L_{2}$. Although, formally, calculations are quite direct,
they are rather lengthy. And, what is surprising, the quantities in terms of
tilted ones can be expressed in terms of original characteristics without a
tilde in a rather simple form.

Omitting details, we list the results.

\subsection{Scenario I}

\begin{equation}
L_{2}=\frac{L_{0}b_{2}}{2m_{0}^{2}}-\frac{\sqrt{d}}{2m_{0}^{2}\sqrt{%
E_{0}^{2}+g_{00}m^{2}}}(L_{0}E_{0}+g_{\phi }\omega m_{0}^{2})\text{,}
\label{L2I}
\end{equation}%
\begin{equation}
L_{1}=\frac{L_{0}b_{1}}{2m_{0}^{2}}+\frac{\sqrt{d}}{2m_{0}^{2}\sqrt{%
E_{0}^{2}+g_{00}m^{2}}}(L_{0}E_{0}+g_{\phi }\omega m_{0}^{2}).  \label{L1I}
\end{equation}%
\begin{equation}
E_{2}=E_{0}\frac{b_{2}}{2m_{0}^{2}}-\frac{\sqrt{d}\sqrt{%
m_{0}^{2}g_{00}+E_{0}^{2}}}{2m_{0}^{2}}  \label{e2I}
\end{equation}%
\begin{equation}
E_{1}=E_{0}\frac{b_{1}}{2m_{0}^{2}}+\frac{\sqrt{d}\sqrt{%
m_{0}^{2}g_{00}+E_{0}^{2}}}{2m_{0}^{2}}  \label{e1I}
\end{equation}%
\begin{equation}
P_{2}=\frac{P_{0}}{2m_{0}^{2}\sqrt{E_{0}^{2}+g_{00}m_{0}^{2}}}(E_{0}\sqrt{d}%
-b_{2}\sqrt{E_{0}^{2}+g_{00}m_{0}^{2}})\text{,}  \label{P2f}
\end{equation}%
\begin{equation}
P_{1}=\frac{P_{0}}{2m_{0}^{2}\sqrt{E_{0}^{2}+g_{00}m_{0}^{2}}}(b_{1}\sqrt{%
E_{0}^{2}+g_{00}m_{0}^{2}}+E_{0}\sqrt{d}).  \label{P1f}
\end{equation}

\subsection{Scenario II}

\begin{equation}
L_{1}=\frac{L_{0}}{2m_{0}^{2}}b_{1}+\frac{\sqrt{d}}{2m_{0}^{2}\sqrt{%
E_{0}^{2}+m_{0}^{2}g_{00}}}(L_{0}E_{0}+g_{\phi }\omega m_{0}^{2})
\label{L2II}
\end{equation}

\begin{equation}
L_{2}=\frac{L_{0}}{2m_{0}^{2}}b_{2}-\frac{\sqrt{d}}{2m_{0}^{2}\sqrt{%
E_{0}^{2}+m_{0}^{2}g_{00}}}(L_{0}E_{0}+g_{\phi }\omega m_{0}^{2})
\label{L1II}
\end{equation}

\begin{equation}
E_{1}=\frac{E_{0}b_{1}}{2m_{0}^{2}}+\frac{\sqrt{d}}{2m_{0}^{2}}\sqrt{%
E_{0}^{2}+m_{0}^{2}g_{00}},  \label{e2II}
\end{equation}%
\begin{equation}
E_{2}=\frac{b_{2}E_{0}}{2m_{0}^{2}}-\frac{\sqrt{d}}{2m_{0}^{2}}\sqrt{%
E_{0}^{2}+m_{0}^{2}g_{00}}.  \label{e1II}
\end{equation}%
\begin{equation}
P_{1}=\frac{P_{0}}{2m_{0}^{2}\sqrt{E_{0}^{2}+g_{00}m^{2}}}(b_{1}\sqrt{%
E_{0}^{2}+g_{00}m_{0}^{2}}+E_{0}\sqrt{d})\text{,}  \label{P2II}
\end{equation}%
\begin{equation}
P_{2}=\frac{P_{0}}{2m_{0}^{2}\sqrt{E_{0}^{2}+g_{00}m^{2}}}(b_{2}\sqrt{%
E_{0}^{2}+g_{00}m_{0}^{2}}-E_{0}\sqrt{d}).  \label{P1II}
\end{equation}

In this scenario, we can interchange labels 1 and 2 since each of the two
particles falls in a black hole.

Additionally, the condition 
\begin{equation}
m_{0}\geq m_{1}+m_{2}  \label{m012}
\end{equation}%
should be fulfilled.

In the nonrelativistic case, the mass is conserved, $m_{0}=m_{1}+m_{2}$.
However, for relativistic particles, inequality $m_{0}>m_{1}+m_{2}$ is
possible due to the contribution of kinetic energy. In particular, decay to
particles with negligible masses of particles 1 and 2 is allowed. It is
worth mentioning that eq. (\ref{m012}) is not an additional assumption but
can be derived from equations of motions - see eqs. (21) - (25) in \cite%
{centr}.

\section{Properties of scenarios\label{prop}}

\subsection{Scenario I}

Let us discuss scenario I firstly. We remind a reader that $P$ is an
absolute value, so it is necessary that $P\geq 0$. As $d\leq b_{1,2}$, it is
clear that inside the ergoregion, where $g_{00}>0$, inequality $P_{2}>0$ in (%
\ref{P2f}) cannot be satisfied. Therefore, scenario I is not realized there
at all. Particle 2 is drifted by flow of space, so particle 2 is ejected
inward in the frame comoving with particle 0, $\dot{r}_{2}<0$ anyway, and
only scenario II remains. It does not mean, of course, that particle 2
cannot reach infinity. Instead, it only means that after decay it moves
towards the horizon. In principle, it can bounce back after reflection from
the potential barrier and escape to infinity. This depends on concrete
properties of the metric.

On the first glance, the impossibility of scenario I inside the ergoregion
looks paradoxical but it can be explained as follows. If two particles move
along the same line tangent to the trajectory (in the same or opposite
directions), they must have $d\phi /dr$ of the same sign. Let $dr_{1}<0$ and 
$dr_{2}>0$. Then, $d\phi _{1}$ and $d\phi _{2}$ must have the opposite
signs. However, inside the ergoregion the sign of the angular velocity is
fixed, particles rotate in the same direction as a black hole due to strong
frame-dragging. Therefore, $d\phi _{1}$ and $d\phi _{2}$ have the same
signs, so scenario I in the ergoregion is impossible.

Outside the ergoregion $g_{00}=-\left\vert g_{00}\right\vert <0$, and
scenario I is possible, provided (\ref{P2f}) is positive, so%
\begin{equation}
E_{0}<\frac{b_{2}\sqrt{\left\vert g_{00}\right\vert }}{2m_{2}}\text{.}
\label{en1}
\end{equation}%
However, the Penrose process is impossible as it should be outside the
ergoregion, since $0<E_{2}<E_{0}$, $0<E_{1}<E_{0}$. Decay occurs without
energy extraction from a black hole.

\subsection{Scenario II}

This type of decay is possible outside the ergoregion ($g_{00}<0$), provided%
\begin{equation}
E_{0}>\frac{b_{2}\sqrt{\left\vert g_{00}\right\vert }}{2m_{2}}\text{.}
\end{equation}

Inside the ergoregion ($g_{00}>0$), this scenario is always possible. Can
the PP be possible as well? It is seen from (\ref{e2II}), (\ref{e1II}) that $%
E_{1}>0$. If we require $E_{2}<0$, this entails%
\begin{equation}
E_{0}<\frac{\sqrt{d}}{2m_{2}}\sqrt{g_{00}}\text{.}  \label{2pp}
\end{equation}

However, in this scenario both particles (at least, immediately after decay)
move towards a black hole. In principle, particle 1 can bounce back from the
potential barrier and escape to infinity. Such a scenario is model-dependent
and is beyond the scope of the present paper.

\subsection{Inverse scenarios}

In a similar way, we can consider the cases when particle 0 moves with $%
\sigma _{0}=+1$ (in the outward direction). For example, it can arrive from
infinity, bounce back in the turning point and keep moving in the outward
direction. Then, the conservation of radial momentum gives us

\begin{equation}
P_{0}=P_{1}+\sigma _{2}P_{2}
\end{equation}%
If $\sigma _{2}=+1$, both particles move outward. In doing so, the formulas
for $E_{i}$, $P_{i}$ and $L_{i}$ ($i=1,2$) coincide with those for scenario
II. Particle 1 can, in principle, move directly to infinity. Particle 2 can
have $E_{2}<0$, provided (\ref{2pp}) holds in the ergoregion. However, it is
inevitable that in this case particle 2 bounces back and moves further
towards a black hole in agreement with general properties of trajections
with negative energy \cite{gpneg}.

Scenario with $\sigma _{2}=-1$ is similar to scenario I.

\section{Maximization of energy of debris\label{maxim}}

We can look at the problem from a somewhat different viewpoint. For fixed $%
L_{0}\,$, $E_{0}$ and given masses $m_{0}$, $m_{1}$, $m_{2}$ eqs. (\ref{e2}%
), (\ref{e1}) leave 1 free parameter, say \thinspace $L_{1}$. Once it is
fixed, one obtains $L_{2}=L_{0}-L_{1}$ and $E_{1}$, $E_{2}$ from
aforementioned equations and the definition (\ref{x}). If $L_{1}$ is allowed
to vary, we have energies as functions of $L_{1}$: $E_{1}=E_{1}(L_{1})$ and $%
E_{2}=E_{2}(L_{1})$. Let us require that the quantity $E_{1}$ reach the
maximum with respect to $L_{1}$. Then,%
\begin{equation}
\frac{\partial E_{1}}{\partial L_{1}}=0.
\end{equation}

As $E_{2}=E_{0}-E_{1}$, for $E_{2}$ we will have minimum. Taking into
account general formulas (\ref{e1}), (\ref{e2}), one obtains 
\begin{equation}
-\delta \sqrt{\tilde{d}}(\omega \tilde{m}_{0}^{2}+\frac{L_{0}X_{0}}{g_{\phi }%
})=\frac{P_{0}}{g_{\phi }}(2L_{2}m_{0}^{2}-b_{2}L_{0}).
\end{equation}

After some manipulations, one can show easily that its solution gives us (%
\ref{L2I}), (\ref{L1I}) or (\ref{L2II}), (\ref{L1II}).

Thus both approaches are equivalent.

\section{Comparison with the Wald approach\label{wald}}

There is one more approach. For decay of \ particles, general inequalities
were derived by Wald \cite{wald}:%
\begin{equation}
E_{\min }\leq E\leq E_{\max }\text{,}
\end{equation}%
where%
\begin{equation}
E_{\max }=\gamma _{2}\frac{m_{2}}{m_{0}}(E_{0}+v_{2}\sqrt{E_{0}^{2}+m_{0}^{2}%
})=E_{2}\text{,}  \label{emax}
\end{equation}%
\begin{equation}
E_{\min }=\gamma _{1}\frac{m_{1}}{m_{0}}(E_{0}-v_{1}\sqrt{E_{0}^{2}+m_{0}^{2}%
})=E_{1}\text{.}  \label{emin}
\end{equation}

Here, $v_{2}$ is the velocity of particle 2 with respect to particle 0. In
other words, this is a relative velocity. Here, $\gamma _{2}=\frac{1}{\sqrt{%
1-v_{2}^{2}}}$ is the corresponding Lorentz gamma factor. Similar quantities
and notations stand for particle 1.

Comparing (\ref{e1II}), (\ref{e2II}) and (\ref{emax}), (\ref{emin}) we see
that both expressions coincide, if we identify%
\begin{equation}
v_{2}=\frac{\sqrt{d}}{b_{2}}\text{, }\gamma _{2}=\frac{b_{2}}{2m_{0}m_{2}}%
\text{.}  \label{vd}
\end{equation}

In a similar way,%
\begin{equation}
v_{1}=\frac{\sqrt{d}}{b_{1}}\text{, }\gamma _{1}=\frac{b_{1}}{2m_{0}m_{1}}.
\label{v1d}
\end{equation}

This is not the end of story. We must prove that this is not formal
coincidence, but $v$ and $\gamma $ defined according to (\ref{vd}) do have
the kinematic meaning of a relative velocity and Lorentz gamma factor. To
this end, we can calculate the relative gamma-factor between particle 0 and
2: $\gamma _{2}=-u_{\mu (0)}u^{\mu (2)}$.

Using the explicit expressions (\ref{u}) for each of particles 0 and 2, one
can show that%
\begin{equation}
\gamma _{2}=\frac{X_{0}X_{2}-\sigma P_{0}P_{2}}{N^{2}}-\frac{L_{2}L_{0}}{%
g_{\phi }}\text{,}
\end{equation}%
where $\sigma =-1$ for scenario I and $\sigma =+1$ for scenario II.

Using (\ref{L1II}) - (\ref{P1II}), we obtain after some calculations that $%
\gamma _{2}$ (hence, also $v_{2}$) does indeed coincide with (\ref{vd}).
Thus two ways of calculations (from equations of motion or by comparing two
expressions ((\ref{e1II}), (\ref{e2II}) and (\ref{emax}), (\ref{emin})) give
the same result. The similar statement holds for particle 1. Thus $v_{1}$
and $v_{2}$ depend on the relationship between masses only.

\section{Decay in turning point\label{tp}}

There exists a special type of scenario, when decay occurs in the turning
point for all three particles, $P_{i}=0$ for $i=0,1,2$. For brevity, we will
call it scenario TP. The condition (\ref{eq}) loses its sense in this case
and turns into an empty identity. Therefore, we must consider the problem
all over again.

Then, according to (\ref{P}), in the point of decay 
\begin{equation}
X_{i}=\tilde{m}_{i}N\text{,}  \label{xt}
\end{equation}%
$i=0,1,2$. It is clear from (\ref{x12}) that this requires 
\begin{equation}
\tilde{m}_{0}=\tilde{m}_{1}+\tilde{m}_{2}\text{.}  \label{m12}
\end{equation}%
Then, it follows from (\ref{P}), (\ref{del}), (\ref{b2t}), (\ref{D}) that $%
\tilde{b}_{1}=2\tilde{m}_{1}\tilde{m}_{0}$, $\tilde{b}_{2}=2\tilde{m}_{0}%
\tilde{m}_{2}$, $\tilde{d}=0$. Then, we have from (\ref{xt}), (\ref{m12})
that%
\begin{equation}
X_{1}=\frac{X_{0}\tilde{m}_{1}}{\tilde{m}_{0}}\text{, }X_{2}=\frac{X_{0}%
\tilde{m}_{2}}{\tilde{m}_{0}}\text{.}  \label{Xm0}
\end{equation}

If the process occurs in the egorsphere, $g_{00}>0$ that is equivalent to 
\begin{equation}
N<\omega \sqrt{g_{\phi }}\text{.}  \label{erg}
\end{equation}

Eq. (\ref{012}) for particle 0 gives us the quadratic equation for $L_{0}$.
In general, it has two roots: 
\begin{equation}
L_{0}=E_{0}\omega \frac{g_{\phi }}{g_{00}}+\frac{N\sqrt{g_{\phi }}}{g_{00}}%
\sqrt{E_{0}^{2}+m_{0}^{2}g_{00}}  \label{L0+}
\end{equation}%
or

\begin{equation}
L_{0}=E_{0}\omega \frac{g_{\phi }}{g_{00}}-\frac{N\sqrt{g_{\phi }}}{g_{00}}%
\sqrt{E_{0}^{2}+m_{0}^{2}g_{00}}\text{.}  \label{L}
\end{equation}%
Then,%
\begin{equation}
X_{0}=\tilde{m}_{0}N=-\frac{E_{0}N^{2}}{g_{00}}\pm \frac{\omega N\sqrt{%
g_{\phi }}}{g_{00}}\sqrt{E_{0}^{2}+m_{0}^{2}g_{00}}\text{.}  \label{x0}
\end{equation}%
However, in the ergoregion $g_{00}>0$, so only root (\ref{L}) is compatible
with (\ref{ft}). Then, one should take the upper sign in (\ref{x0}). Outside
the ergoregion, both signs in (\ref{x0}) are suitable. In doing so, sign
"minus" in (\ref{x0} corresponds to (\ref{L0+}).

The expressions for $L_{1},L_{2}$ follow from (\ref{m12}) and (\ref{L0}).
Then, after some calculations that include taking the square, one obtains%
\begin{equation}
L_{1}=\frac{b_{1}}{2m_{0}^{2}}L_{0}\pm \frac{\sqrt{g_{\phi }d}\tilde{m}_{0}}{%
2m_{0}^{2}},  \label{1tp}
\end{equation}%
\begin{equation}
L_{2}=\frac{b_{2}}{2m_{0}^{2}}L_{0}\mp \frac{\sqrt{g_{\phi }d}\tilde{m}_{0}}{%
2m_{0}^{2}},  \label{l2tp}
\end{equation}%
\begin{equation}
\tilde{m}_{1}=\frac{b_{1}\sqrt{L_{0}^{2}+g_{\phi }m_{0}^{2}}\pm \sqrt{d}L_{0}%
}{2m_{0}^{2}\sqrt{g_{\phi }}}=\frac{b_{1}}{2m_{0}^{2}}\tilde{m}_{0}\pm \frac{%
\sqrt{d}L_{0}}{2m_{0}^{2}\sqrt{g_{\phi }}}
\end{equation}%
\begin{equation}
\tilde{m}_{2}=\frac{b_{2}}{2m_{0}^{2}}\tilde{m}_{0}\mp \frac{\sqrt{d}L_{0}}{%
2m_{0}^{2}\sqrt{g_{\phi }}}.
\end{equation}%
It is seen that (\ref{m12}) is satisfied as it should be.

Using also the definition of $X$ and (\ref{L0+}), (\ref{L}), we have after
some algebraic manipulations that%
\begin{equation}
E_{1}=\frac{b_{1}}{2m_{0}^{2}}E_{0}\pm \frac{\sqrt{d}}{2m_{0}^{2}}\sqrt{%
E_{0}^{2}+m_{0}^{2}g_{00}}\text{,}  \label{e1t}
\end{equation}%
\begin{equation}
E_{2}=\frac{b_{2}}{2m_{0}^{2}}E_{0}\mp \frac{\sqrt{d}}{2m_{0}^{2}}\sqrt{%
E_{0}^{2}+m_{0}^{2}g_{00}}  \label{e2t}
\end{equation}%
that agrees with (\ref{e1I}), (\ref{e2I}) or (\ref{e1II}), (\ref{e2II}).

It is essential that signs in (\ref{e1t}), (\ref{e2t}) coincide with those
in (\ref{1tp}), (\ref{l2tp}). Now, particles 1 and 2 can be considered on
equal footing. However, the subtle point is that signs in (\ref{1tp}),(\ref%
{l2tp}) should agree with signs for the energy.

Further, we can substitute $L_{0}$ and express $L_{1,2}$ in terms of energy $%
E_{0}$: 
\begin{equation}
L_{2}=E_{0}\frac{b_{2}\omega g_{\phi }\mp N\sqrt{d}\sqrt{g_{\phi }}}{%
2m_{0}^{2}g_{00}}+\frac{\sqrt{E_{0}^{2}+m_{0}^{2}g_{00}}}{2m_{0}^{2}g_{00}}%
(-b_{2}N\sqrt{g_{\phi }}\pm \sqrt{d}g_{\phi }\omega ),
\end{equation}%
\begin{equation}
L_{1}=E_{0}\frac{b_{1}\omega g_{\phi }\pm N\sqrt{d}\sqrt{g_{\phi }}}{%
2m_{0}^{2}g_{00}}+\frac{\sqrt{E_{0}^{2}+m_{0}^{2}g_{00}}}{2m_{0}^{2}g_{00}}%
(-b_{1}N\sqrt{g_{\phi }}\mp \sqrt{d}g_{\phi }\omega ).
\end{equation}

In the case under discussion, classification of scenarios that includes I
and II is not valid any longer since all $P_{i}=0$ by definition. If we are
interested in the question, whether or not particle (say, 2) escapes after
decay, one is led to condition $h_{2}>0$ where $h_{2}$ is given by (\ref{h12}%
).

\section{Penrose process and signs of angular momenta for decay in turning
point\label{signs}}

For decay in a generic point we discussed above some features of the process
connected with radial components of particle velocities. In scenario TP, by
definition, radial components of momenta $P_{i}=0$ for $i=0,1,2$. Meanwhile,
here it is of some interest to elucidate the properties of the scenario that
are connected with angular motion. This concerns both angular velocities and
momenta.

\subsection{Angular velocities}

It is instructive to find $\Omega _{0}.$ Using eq. (\ref{ome}) and (\ref{L})
we obtain%
\begin{equation}
\Omega _{0}=\omega +\frac{\varepsilon _{0}N}{\sqrt{g_{\phi }}\sqrt{%
\varepsilon _{0}^{2}+g_{00}}}\text{,}  \label{o0}
\end{equation}%
where $\varepsilon =\frac{E}{m}.$

It follows from (\ref{ome}) and (\ref{Xm0}) that

\begin{equation}
\Omega _{1,2}=\omega +\frac{LN}{\sqrt{g_{\phi }}\sqrt{%
L_{1,2}^{2}+m_{1,2}^{2}g_{\phi }}}.  \label{om12}
\end{equation}

As is well known, the condition that the interval is time-like or light-like
gives for a particle with $r=const$%
\begin{equation}
\Omega _{-}\leq \Omega \leq \Omega _{+}\text{,}  \label{om+-}
\end{equation}%
where

\begin{equation}
\Omega _{+}=\omega +\frac{N}{\sqrt{g_{\phi }}}
\end{equation}%
is the maximum possible value for a particle, 
\begin{equation}
\Omega _{-}=\omega -\frac{N}{\sqrt{g_{\phi }}}
\end{equation}%
is the minimum one. Inside the ergoregion, $\Omega _{-}>0$ and this causes
orbital motion of a particle.

If $m_{i}=0$, (\ref{om12}) gives us%
\begin{equation}
\Omega _{1,2}=\omega +\frac{\sigma _{1,2}N}{\sqrt{g_{\phi }}}\text{,}
\end{equation}%
where $\sigma _{i}=\pm 1$ depending on \ the sign of $L_{i}$. Is it possible
to have the same signs for both massless particles? Below, we show in a
rather general setting that the answer is "no".

If $m_{1}=m_{2}=0$, we can derive some more features of angular momenta. In
this case, $b_{1}=b_{2}=m_{0}^{2}=\sqrt{d}$. Then, it follows from (\ref{L2I}%
), (\ref{L1I}) or (\ref{L2II}), (\ref{L1II}) that $signL_{1}L_{2}=signY$,
where%
\begin{equation}
Y=L_{0}^{2}-\frac{1}{(E_{0}^{2}+m_{0}^{2}g_{00})}(L_{0}E_{0}+g_{\phi }\omega
m_{0}^{2})^{2}.  \label{Yang}
\end{equation}

For scenario TP under discussion, with (\ref{L}) taken into account,%
\begin{equation}
Y=-m_{0}^{2}g_{\phi }<0\text{.}
\end{equation}

Thus $L_{1}L_{2}<0$. Correspondingly, for one of particles $\Omega =\Omega
_{+}$, for another one $\Omega =\Omega _{-}$. Both particles appear with
opposite angular momenta and realize the maximum and minimum values of $%
\Omega $.

For massive particles, a general inequality (\ref{om+-}) holds. What can be
said about the relative signs of angular momenta in this case? Let particle
0 decay inside the ergoregion and the Penrose process occur. Then, for one
of particles (say, particle 1) $E_{1}<0$. We must have $L_{1}<0$ according
to the forward-in-time condition (\ref{ft}). Now, we are going to elucidate,
whether it is possible to have $L_{2}<0$ as well.

Let us assume that indeed $L_{2}<0$. Then, $L_{0}=L_{1}+L_{2}=-\left\vert
L_{0}\right\vert <0$ as well. According to (\ref{L1II}), we have the
following general condition on $\left\vert L_{0}\right\vert $:%
\begin{equation}
\left\vert L_{0}\right\vert >\frac{\sqrt{d}g_{\phi }\omega m_{0}^{2}}{\sqrt{%
E_{0}^{2}+m_{0}^{2}g_{00}}}(b_{2}+\sqrt{d}\frac{E_{0}}{\sqrt{%
E_{0}^{2}+m_{0}^{2}g_{00}}})^{-1}.  \label{le}
\end{equation}

In scenario TP, further restrictions can be obtained. Then, \ $L_{0}$ is not
independent quantity but is given by (\ref{L}), so $L_{0}<0$ entails 
\begin{equation}
N\sqrt{E_{0}^{2}+m_{0}^{2}g_{00}}>E_{0}\omega \sqrt{g_{\phi }}.
\end{equation}

In combination with (\ref{erg}) this gives us%
\begin{equation}
N_{0}\equiv \omega \sqrt{g_{\phi }}\frac{E_{0}}{\sqrt{%
E_{0}^{2}+m_{0}^{2}g_{00}}}<N<\omega \sqrt{g_{\phi }}.  \label{N}
\end{equation}

Also, we obtain from (\ref{le}) and (\ref{L}) that%
\begin{equation}
N>N_{1}\equiv \frac{1}{\sqrt{E_{0}^{2}+m_{0}^{2}g_{00}}}(E_{0}\omega \sqrt{%
g_{\phi }}+\frac{g_{00}\sqrt{d}\omega \sqrt{g_{\phi }}m_{0}^{2}}{b_{2}\sqrt{%
E_{0}^{2}+m_{0}^{2}g_{00}}+E_{0}\sqrt{d}})  \label{N1}
\end{equation}%
Obviously, $N_{1}>N_{0}$, so (\ref{N1})\ is a more tight condition that $%
N>N_{0}$. Thus for $L_{2}<0$ we must have%
\begin{equation}
N_{1}<N<\omega \sqrt{g_{\phi }}\text{,}  \label{strip}
\end{equation}%
whence $N_{1}<\omega \sqrt{g_{\phi }}$. It is easy to check that this \
holds for any $m_{2}\neq 0$. Then, only in the region (\ref{strip}) the
angular momentum $L_{2}<0$.

Meanwhile, if particle 2 is massless, $m_{2}=0$, this strip degenerates to
the point, so actually $L_{2}<0$ is impossible in accordance with what is
written above.

\section{Threshold for the Penrose process\label{thr}}

We are interested in the situation when $E_{2}>E_{0}$, so that the
efficiency $\eta \equiv \frac{E_{2}}{E_{0}}>1$. Let us consider scenario TM
assuming that particle 2 moves after decay in the outward direction. Then,
it follows from (\ref{e2t}) where sign "plus" is taken that

\begin{equation}
b_{2}+\sqrt{d}\sqrt{1+\frac{g_{00}}{\varepsilon _{0}^{2}}}>2m_{0}^{2}\text{,}
\end{equation}%
whence%
\begin{equation}
v_{2}>v_{2\min }=\frac{b_{1}}{b_{2}}\frac{1}{\sqrt{1+\frac{\left(
g_{00}\right) _{H}}{\varepsilon ^{2}}}},  \label{v2}
\end{equation}%
where we used (\ref{vd}) and (\ref{b}), (\ref{b2}).

In a similar way, condition $E_{1}<0$ gives us%
\begin{equation}
v_{1}>v_{1\min }=\frac{1}{\sqrt{1+\frac{g_{00}}{\varepsilon _{0}^{2}}}}=%
\frac{\left( v_{2}\right) _{\min }b_{2}}{b_{1}}=\frac{1}{\sqrt{1+\frac{g_{00}%
}{\varepsilon _{0}^{2}}}}\text{.}  \label{v1}
\end{equation}%
Both conditions are equivalent to each other, as it should be, according to (%
\ref{vd}), (\ref{v1d}).

For the Kerr metric, taking $g_{00}=1$ on the extremal horizon in the
equatorial plane and choosing $\varepsilon _{0}=1$, we obtain $v_{1\min }=%
\frac{1}{\sqrt{2}}$. For circular orbits around the near-extremal horizon,
we can put $\varepsilon _{0}=\frac{1}{\sqrt{3}}$, then $v_{1\min }=\frac{1}{2%
}$. These results agree with \cite{72}.

Eqs. (\ref{v2}) and (\ref{v1}) with (\ref{vd}), (\ref{v1d}) taken into
account can be rewritten as%
\begin{equation}
b_{1}>2m_{0}m_{1}\gamma _{1\min }\text{,}
\end{equation}%
\begin{equation}
m_{0}^{2}-2m_{0}m_{1}\gamma _{1\min }+m_{1}^{2}-m_{2}^{2}>0\text{,}
\end{equation}%
where $\gamma _{1\min }=\frac{1}{\sqrt{1-\left( v_{1}\right) _{\min }^{2}}}=%
\frac{\sqrt{\varepsilon _{0}^{2}+g_{00}}}{\sqrt{g_{00}}}$, whence%
\begin{equation}
m_{0}>m_{+},\text{ }  \label{m0}
\end{equation}%
or 
\begin{equation}
m_{0}<m_{-}\text{,}  \label{mmin}
\end{equation}%
where%
\begin{equation}
m_{+}=\frac{m_{1}\sqrt{\varepsilon _{0}^{2}+g_{00}}+\sqrt{%
m_{1}^{2}\varepsilon _{0}^{2}+m_{2}^{2}g_{00}}}{\sqrt{g_{00}}},  \label{m+}
\end{equation}%
\begin{equation}
m_{-}=\frac{m_{1}\sqrt{\varepsilon _{0}^{2}+g_{00}}-\sqrt{%
m_{1}^{2}\varepsilon _{0}^{2}+m_{2}^{2}g_{00}}}{\sqrt{g_{00}}}=\frac{%
(m_{1}^{2}-m_{2}^{2})\sqrt{g_{00}}}{m_{1}\sqrt{\varepsilon _{0}^{2}+g_{00}}+%
\sqrt{m_{1}^{2}\varepsilon _{0}^{2}+m_{2}^{2}g_{00}}}\text{.}
\end{equation}

It is clear that $m_{+}>m_{1}+m_{2}$, so (\ref{m0}) is a more tight
condition than (\ref{m012}) due to the account for contribution of kinetic
energy of particles 1 and 2. Also, it is easy to show that (\ref{m012}) is
inconsistent with (\ref{mmin}). Therefore, we are left with (\ref{m0}) only.

If $m_{1}=m_{2}\equiv m$, we must have 
\begin{equation}
m_{0}>\frac{2m}{\sqrt{g_{00}}}\sqrt{\varepsilon _{0}^{2}+g_{00}}\text{.}
\end{equation}

\subsection{Circle orbits}

Further information about the scenario under discussion can be found if we
consider not simply turning points but assume that the corresponding point
lies on the circle orbit. General estimates for "dirty" (surrounded by
matter) extremal black holes according to eq. (66) of \cite{near} give us
for near-horizon circular orbits

\begin{equation}
\varepsilon =\frac{\omega _{H}\sqrt{g_{\phi }}}{\sqrt{B_{1}^{2}g_{\phi }-1}},
\end{equation}%
where $B_{1}$ is the coefficient in the near-horizon expansion%
\begin{equation}
\omega =\omega _{H}-B_{1}N+...
\end{equation}

For the extremal Kerr-Newman%
\begin{equation}
\varepsilon =\frac{a}{M\sqrt{4\frac{a^{2}}{M^{2}}-1}}\text{,}
\end{equation}%
\begin{equation}
B_{1}=\frac{2a}{a^{2}+M^{2}}\text{,}
\end{equation}%
\begin{equation}
\left( g_{00}\right) _{H}=\left( \omega ^{2}g_{\phi }\right) _{H}=\frac{a^{2}%
}{M^{2}},
\end{equation}%
where $M$ is the mass of a black hole, $a$ being its angular momentum. Then,%
\begin{equation}
v_{1}=\frac{1}{2}\frac{M}{a}\text{, }v_{2}=\frac{1}{2}\frac{M}{a}\frac{b_{1}%
}{b_{2}}.
\end{equation}

In accordance with \cite{wald} and eq. A15 of \cite{72} and (see also page
373 of \cite{chandra}), $v_{1}>\frac{1}{2}$.

\section{Rotating black holes: efficiency in particular cases\label{efrot}}

Now, we are going to analyze the efficiency of the process depending on the
mass of particles participating in decay, especially we will discuss the
near-horizon limit. We assume that it is particle 2 that has positive energy
and escapes - either immediately (in scenario TP) or later, after reflection
from the potential barrier (in scenario AP). In doing so, its energy is
described by eq. (\ref{e2t}) with sigh "plus".

Then, the efficiency

\begin{equation}
\eta =\frac{E_{2}}{E_{0}}=\frac{b_{2}}{2m_{0}^{2}}+\frac{\sqrt{d}}{2m_{0}^{2}%
}\sqrt{1+\frac{g_{00}}{\varepsilon _{0}^{2}}}\text{,}  \label{ef}
\end{equation}%
$\varepsilon =\frac{E}{m}$.

It is remarkable that the metric enters the efficiency through the component 
$g_{00}$ only. This simple circumstance has some important consequences. If $%
g_{00}$ is \ a monotonically decreasing function of the radial coordinate
(like for the Kerr metric), the maximum of efficiency is reached when the
turning point is as close to the horizon as possible. Let us consider it in
more detail.

\subsubsection{Decay on the horizon}

\begin{equation}
\eta =\frac{E_{2}}{E_{0}}=\frac{b_{2}}{2m_{0}^{2}}+\frac{\sqrt{d}}{2m_{0}^{2}%
}\sqrt{1+\frac{\left( g_{\phi }\omega ^{2}\right) _{H}}{\varepsilon ^{2}}}.
\label{eff}
\end{equation}%
Hereafter, subscript "H" means that a corresponding quantity is calculated
on the horizon.

For given masses $m_{0}$, $m_{1}$ and $m_{2}$ and $E_{0}\geq m_{0}$, the
maximum of (\ref{ef}) is reached for the minimum possible value $\varepsilon
=1$, if a particle 0 falls from infinity.

Let $m_{0}$ be fixed. Obviously,

\begin{equation}
m_{0}\geq m_{1}+m_{2}.  \label{m0>}
\end{equation}

Let us consider different particular cases.

$m_{0}=m_{1}+m_{2}$.

Then, $d=0$,%
\begin{equation}
\eta =\frac{m_{2}}{m_{0}}<1\text{,}
\end{equation}%
so that the energy extraction is impossible.

$m_{2}=0$.

Then, $d=b_{2}^{2}$,%
\begin{equation}
\eta =\frac{b_{2}}{2m_{0}^{2}}(1+\sqrt{1+\frac{g_{00}}{\varepsilon ^{2}}}).
\end{equation}

If also $m_{1}=0$ (two photons)%
\begin{equation}
\eta =\frac{1+\sqrt{1+\frac{g_{00}}{\varepsilon ^{2}}}}{2}.  \label{max}
\end{equation}

If particle 0 comes from infinity, the maximum is achieved, if decay occurs
on the horizon and $\varepsilon =1$, then%
\begin{equation}
\eta =\frac{1+\sqrt{1+g_{00}}}{2}.
\end{equation}

This eq. coincides with eq. (7) of \cite{j} if one puts there $g_{t\phi
}=-\omega g_{\phi }$ and takes into account that on the horizon $N=0$, so $%
g_{00}=\omega ^{2}g_{\phi }.$

On the extremal horizon of the Kerr metric $g_{00}=+1$ and \cite{shn}%
\begin{equation}
\eta =\frac{1+\sqrt{2}}{2}\text{, }\eta -1=\frac{\sqrt{2}-1}{2}.
\end{equation}

For the extremal Kerr circle orbit, $\varepsilon _{0}=\frac{1}{\sqrt{3}}$, $%
g_{00}=1$, for decay to two photons we have%
\begin{equation}
\eta =\frac{1+\sqrt{1+3g_{00}}}{2}.
\end{equation}

On the near-horizon orbit $g_{00}=1$, $\eta =\frac{3}{2}$.

Let $m_{1}=m_{2}=m$. Then, $b_{2}=b_{1}=m_{0}^{2}$, $%
d=m_{0}^{2}(m_{0}^{2}-4m^{2})$. Putting also $\varepsilon =1$ and for the
horizon taking again $g_{00}=+1$, we obtain%
\begin{equation}
\eta =\frac{1}{2}+\frac{\sqrt{1-4\frac{m_{2}^{2}}{m_{0}^{2}}}\sqrt{2}}{2}.
\end{equation}%
that coincides, for example, with eq. 3.30 of \cite{pani}. Then, $\eta \geq
1 $, provided%
\begin{equation}
\text{ }\frac{m_{2}}{m_{0}}\leq \frac{1}{2\sqrt{2}}\approx 0.35.
\end{equation}

If $\varepsilon \gg 1$,%
\begin{equation}
\eta _{\max }\approx 1+\frac{1}{4}\frac{g_{00}}{\varepsilon ^{2}}.
\end{equation}

If $\varepsilon \rightarrow 0$, so particle 0 starts its motion not from
infinity but from the point that is already close to the horizon from the
very beginning, then%
\begin{equation}
\eta _{\max }\approx \frac{\sqrt{g_{00}}}{2\varepsilon }.
\end{equation}

Now we pose the following question. For a given $m_{0}$, how to choose the
masses $m_{1}$, $m_{2}$ if we want to achieve the maximum efficiency?

It follows from (\ref{eff}) that%
\begin{equation}
\eta =\frac{1}{2}(1+x_{2}-x_{1})+\frac{1}{2}\sqrt{%
1-2x_{1}-2x_{2}+(x_{2}-x_{1})^{2}}B\text{,}
\end{equation}%
where $x_{i}=\frac{m_{i}}{m_{0}}\leq 1$, $i=1,2$,%
\begin{equation}
B=\sqrt{1+\frac{\left( g_{\phi }\omega ^{2}\right) _{H}}{\varepsilon _{0}^{2}%
}}.
\end{equation}%
One can calculate%
\begin{equation}
2\left( \frac{\partial \eta }{\partial x_{1}}\right) _{x_{2}}=-1+\frac{%
B(x_{1}-x_{2}-1)}{\sqrt{1-2x_{1}-2x_{2}+(x_{2}-x_{1})^{2}}}<0
\end{equation}%
As each term here is negative, $\left( \frac{\partial \eta }{\partial x_{1}}%
\right) _{x_{2}}<0$. The maximum is reached for the minimum possible $%
x_{1}=0 $. Then, (\ref{eff}) gives us%
\begin{equation}
\eta =\frac{m_{0}^{2}+m_{2}^{2}}{2m_{0}^{2}}+\frac{m_{0}^{2}-m_{2}^{2}}{%
2m_{0}^{2}}B.
\end{equation}

This should be supplemented with the escaping condition which is
model-dependent.

In the horizon limit $N\rightarrow 0$%
\begin{equation}
X_{2}\approx \frac{N\omega \sqrt{g_{\phi }}}{2m_{0}^{2}\left( \omega
^{2}g_{\phi }\right) _{H}}[b_{2}\sqrt{E_{0}^{2}+m_{0}^{2}\left( \omega
^{2}g_{\phi }\right) _{H}}\pm E_{0}\sqrt{d}],
\end{equation}%
\begin{equation}
L_{2}\approx \omega _{H}g_{\phi H}(E_{0}\frac{b_{2}}{2m_{0}^{2}\omega
_{H}^{2}g_{\phi _{H}}}\pm \frac{\sqrt{E_{0}^{2}+m_{0}^{2}\omega
_{H}^{2}g_{\phi _{H}}}}{2m_{0}^{2}\omega _{H}^{2}g_{\phi _{H}}}\sqrt{d}).
\end{equation}

For both signs, eq. (\ref{ft}) is fulfilled.

\begin{equation}
E_{2}\approx \frac{E_{0}}{2m_{0}^{2}}b_{2}+\frac{\sqrt{d}\sqrt{%
E_{0}^{2}+\omega ^{2}g_{\phi H}}}{2m_{0}^{2}}\text{,}
\end{equation}%
\begin{equation}
E_{1}\approx \frac{E_{0}b_{1}}{2m_{0}^{2}}-\frac{\sqrt{d}}{2m_{0}^{2}}\sqrt{%
E_{0}^{2}+\omega ^{2}g_{\phi H}}.
\end{equation}

If, in scenario TP, we want particle 2 to escape, we must impose condition $%
h_{2}>0$ where $h_{2}$ is given by eq. (\ref{h12}). In a general case, in
the horizon limit $N\rightarrow 0$ this condition cannot be fulfilled.
Formally, there is an exception when 
\begin{equation}
m_{2}=0\text{, }L_{2}=0.  \label{m2L2}
\end{equation}%
However, the latter condition is inconsistent with (\ref{l2tp}), for any $%
m_{0}\neq 0$. Actually, this means that the process with escape is
impossible in \ this case.

Meanwhile, this can becomes possible if both particles are ejected not in
the turning point and not parallel to the trajectory of particle 0. Then,
the efficiency is less than its maximum value. But, in this case eq. (\ref%
{m2L2})\ opens a possibility for particle (say, 2) to escape even if decay
occurs near the horizon.

\section{Efficiency: short summary of properties for the rotating space-times%
}

\subsection{Scenario TP}

Let decay happen in the turning point. Then, there are the following
properties. 1) If $g_{00}$ is monotonically increasing function of $r$, the
maximum of $\eta $ is achieved at the horizon. 2) If particle 0 comes from
infinity, the maximum of $\eta $ is reached for $\varepsilon =1$. 3) For a
given $m_{0}$, the maximum of $\eta $ is reached if an escaping particle is
massless. The properties 1) - 2) are mentioned in literature time to time
for different concrete models. Here, we established them in a \
model-independent way. Property 3) seems to be new.

\subsection{Scenario AP}

For decay in the ergoregion, both particles 1 and 2 move immediately after
decay in the same direction as particle 0 (provided decay occurs not in the
turning point). The calculation of efficiency has meaning, only if particle
2 bounces back and eclipse to infinity or if particle 0 moves before decay
in the outward direction.

We would like to stress that our conclusions are not universal. If ejection
is performed not parallel to the direction of motion, they do not apply. In
particular, then particle 2 can escape if $m_{2}=0$, $L_{2}=0$. Thus, there
are cases beyond the scope of the present article (see, e.g. \cite{esc1} - 
\cite{tz}).

\section{Static black holes and neutral particles\label{stat}}

In this section we consider characteristics of particle decay in the static
case. As before, we assume that all particles are neutral. Correspondingly,
the Penrose process is impossible. Nonetheless, such a decay can be of
interest since it models, for example, the processes of ejecting fuel from a
relativistic rocket \cite{rocket} when the Penrose process is absent, so the
efficiency $\eta <1$. There is one more motivation for discussion of this
issue. There exists a version of the Penrose process for static black holes
with electrically charged particles \cite{den}, \cite{ruf}. Therefore, it is
desirable to have formulas for neutral ones for future comparison of results
for the electric Penrose process and verifying corresponding limiting
transitions when all charges tend to zero.

There is a crucial difference between the static and stationary cases in the
sense that, as we saw it, if the process occurs in the ergosphere (that is
absent for static black holes) some scenarios are forbidden. For static
metrics, they are in general allowed. Therefore, decay in the static
background should be considered not just as a particular case of the process
near rotating black holes but, rather, as a separate issue. Now, $\omega =0$%
, $g_{00}=-N^{2}$. As before, we consider the most interesting case when
debris of decay are ejected along the trajectory.

\subsection{Scenario I}

\begin{equation}
L_{2}=\frac{L_{0}}{2m_{0}^{2}}(b_{2}-\frac{\sqrt{d}}{\sqrt{%
E_{0}^{2}-m_{0}^{2}N^{2}}}E_{0})\text{,}  \label{L2stat}
\end{equation}%
\begin{equation}
L_{1}=\frac{L_{0}}{2m_{0}^{2}}(b_{1}+\frac{\sqrt{d}}{\sqrt{%
E_{0}^{2}-m_{0}^{2}N^{2}}}E_{0}).  \label{L1stat}
\end{equation}%
\begin{equation}
E_{2}=E_{0}\frac{b_{2}}{2m_{0}^{2}}-\frac{\sqrt{d}\sqrt{%
E_{0}^{2}-m_{0}^{2}N^{2}}}{2m_{0}^{2}}  \label{e2st}
\end{equation}%
\begin{equation}
E_{1}=E_{0}\frac{b_{1}}{2m_{0}^{2}}+\frac{\sqrt{d}\sqrt{%
E_{0}^{2}-m_{0}^{2}N^{2}}}{2m_{0}^{2}}
\end{equation}%
\begin{equation}
P_{2}=\frac{P_{0}}{2m_{0}^{2}\sqrt{E_{0}^{2}+g_{00}m_{0}^{2}}}(E_{0}\sqrt{d}%
-b_{2}\sqrt{E_{0}^{2}-m_{0}^{2}N^{2}})\text{,}
\end{equation}%
\begin{equation}
P_{1}=\frac{P_{0}}{2m_{0}^{2}\sqrt{E_{0}^{2}+g_{00}m_{0}^{2}}}(b_{1}\sqrt{%
E_{0}^{2}-m_{0}^{2}N^{2}}+E_{0}\sqrt{d}).
\end{equation}%
\begin{equation}
E_{0}<\frac{b_{2}}{2m_{2}}N\text{.}  \label{c1}
\end{equation}

\subsection{Scenario II}

\begin{equation}
L_{1}=\frac{L_{0}}{2m_{0}^{2}}(b_{1}+\frac{\sqrt{d}}{\sqrt{%
E_{0}^{2}-m_{0}^{2}N^{2}}}E_{0})\text{,}
\end{equation}%
\begin{equation}
L_{2}=\frac{L_{0}}{2m_{0}^{2}}(b_{2}-\frac{\sqrt{d}}{\sqrt{%
E_{0}^{2}-m_{0}^{2}N^{2}}}E_{0}).
\end{equation}

\begin{equation}
E_{1}=\frac{E_{0}b_{1}}{2m_{0}^{2}}+\frac{\sqrt{d}}{2m_{0}^{2}}\sqrt{%
E_{0}^{2}-m_{0}^{2}N^{2}}
\end{equation}%
\begin{equation}
E_{2}=\frac{b_{1}E_{0}}{2m_{0}^{2}}-\frac{\sqrt{d}}{2m_{0}^{2}}\sqrt{%
E_{0}^{2}-m_{0}^{2}N^{2}}
\end{equation}%
\begin{equation}
P_{1}=\frac{P_{0}}{2m_{0}^{2}\sqrt{E_{0}^{2}+g_{00}m^{2}}}(b_{2}\sqrt{%
E_{0}^{2}-m_{0}^{2}N^{2}}+E_{0}\sqrt{d})\text{,}
\end{equation}%
\begin{equation}
P_{2}=\frac{P_{0}}{2m_{0}^{2}\sqrt{E_{0}^{2}+g_{00}m^{2}}}(b_{2}\sqrt{%
E_{0}^{2}-m_{0}^{2}N^{2}}-E_{0}\sqrt{d}).
\end{equation}%
\begin{equation}
E_{0}>\frac{b_{2}}{2m_{2}}N\text{.}  \label{c2}
\end{equation}

Here, conditions (\ref{c1}), (\ref{c2}) follow from requirements $P_{1,2}>0$%
. The efficiency in AP, scenario I

\begin{equation}
\eta _{I}=\frac{E_{2}}{E_{0}}=\frac{b_{2}}{2m_{0}^{2}}-\frac{\sqrt{d}}{%
2m_{0}^{2}}\sqrt{1-\frac{N^{2}}{\varepsilon ^{2}}}\text{.}
\end{equation}

As $b_{2}-\sqrt{d}<2m_{0}^{2}$, the efficiency $\eta <1$, since for static
metrics the ergoregion for neutral particles is absent, and the Penrose
process is absent as well.

\subsection{Scenario TP}

it follows from (\ref{L0+}), (\ref{L}) with $\omega =0$ that%
\begin{equation}
L_{0}=\pm \frac{1}{N}\sqrt{g_{\phi }(E_{0}^{2}-m_{0}^{2}N^{2})}
\end{equation}%
can have any sign. {}

For scenario TP the maximum efficiency

\begin{equation}
\eta _{TP}=\frac{E_{2}}{E_{0}}=\frac{b_{2}}{2m_{0}^{2}}+\frac{\sqrt{d}}{%
2m_{0}^{2}}\sqrt{1-\frac{N^{2}}{\varepsilon ^{2}}}\text{.}
\end{equation}

This can also be rewritten as%
\begin{equation}
\eta =\frac{b_{2}}{2m_{0}^{2}}+\frac{\sqrt{d}}{2m_{0}^{2}}\frac{\left\vert
L_{0}\right\vert }{\sqrt{L_{0}^{2}+g_{\phi }m_{0}^{2}}}.
\end{equation}%
As $b_{2}+\sqrt{d}<2m_{0}^{2}$, this efficiency $\eta <1$, as it should be.

Different velocities are related according to the local Lorentz
transformation%
\begin{equation}
V_{2}=\frac{v_{2}-V_{0}}{1-v_{2}V_{0}}\text{,}  \label{v20}
\end{equation}%
where $v_{2}$ is the velocity of particle 2 in the frame comoving with
particle 0 and is given by (\ref{vd}). The velocities $V_{0,2}$ are measured
in the laboratory (static) frame. They can be found from the relation%
\begin{equation}
E=\frac{mN}{\sqrt{1-V^{2}}}\text{.}
\end{equation}%
One can check by straightforward calculations that (\ref{v20}) is indeed
valid.

\subsection{Near-horizon limit}

To elucidate, what can happen to decay near a black hole, let us consider
the near-horizon limit when $r\rightarrow r_{h}$, where $r_{h}$ is the
horizon radius.

Let particle 0 turn into particles 1 and 2. Scenario TP is impossible now
since near the horizon where is no turning point for any particle coming
from infinity. Indeed, according to (\ref{xt}) with $\omega =0$, we would
have $E_{0}=O(N)$. However, for a particle under discussion $E_{0}\geq m_{0}$%
. Let us consider scenario of type AP for a decay in a point which is not
the turning one.

\subsubsection{Scenario I}

Particle 1 falls in a black hole, particle 2 moves away from the horizon. It
follows from (\ref{c1}) that for $m_{2}\neq 0$ the point of decay cannot
approach the horizon arbitrarily close to the horizon since the limit $%
N\rightarrow 0$ is inconsistent with $E_{0}\geq m_{0}$. However, if $m_{2}=0$
it becomes possible. The same is true if $m_{2}=O(N)$ is extremely small.
For simplicity, we assume that $m_{2}=0$ exactly. Then, $b_{2}=\sqrt{d}%
=m_{0}^{2}-m_{1}^{2}$, $b_{1}=m_{0}^{2}+m_{1}^{2}$,%
\begin{equation}
P_{2}=\frac{P_{0}(m_{0}^{2}-m_{1}^{2})}{2m_{0}^{2}}(\frac{E_{0}}{\sqrt{%
E_{0}^{2}-m_{0}^{2}N^{2}}}-1),
\end{equation}%
\begin{equation}
P_{1}=\frac{P_{0}}{2m_{0}^{2}}(b_{1}+\frac{E_{0}b_{2}}{\sqrt{%
E_{0}^{2}-m_{0}^{2}N^{2}}}),
\end{equation}%
\begin{equation}
E_{2}=\left( \frac{b_{2}}{2m_{0}^{2}}\right) (E_{0}-\sqrt{%
E_{0}^{2}-m_{0}^{2}N^{2}}),  \label{e21n}
\end{equation}%
\begin{equation}
\eta =\frac{E_{2}}{E_{0}}=\left( \frac{b_{2}}{2m_{0}^{2}}\right) (1-\sqrt{1-%
\frac{m_{0}^{2}}{E_{0}^{2}}N^{2}}),  \label{efst}
\end{equation}%
\begin{equation}
E_{1}=\left( \frac{b_{1}}{2m_{0}^{2}}\right) (E_{0}+\sqrt{%
E_{0}^{2}-m_{0}^{2}N^{2}}),
\end{equation}%
\begin{equation}
L_{2}=b_{2}\frac{L_{0}}{2m_{0}^{2}}(1-\frac{E_{0}}{\sqrt{%
E_{0}^{2}-m_{0}^{2}N^{2}}})\text{,}
\end{equation}%
\begin{equation}
L_{1}=\frac{L_{0}}{2m_{0}^{2}}(b_{1}+\frac{b_{2}}{\sqrt{%
E_{0}^{2}-m_{0}^{2}N^{2}}}E_{0}).
\end{equation}%
For decay near the horizon, $N\ll 1$ and%
\begin{equation}
E_{2}\approx \frac{m_{0}^{2}-m_{1}^{2}}{4E_{0}}N^{2}\text{,}  \label{e2n}
\end{equation}%
\begin{equation}
E_{1}\approx E_{0}-\frac{(m_{0}^{2}-m_{1}^{2})N^{2}}{4E_{0}}\text{,}
\end{equation}%
\begin{equation}
P_{2}\approx \frac{P_{0}(m_{0}^{2}-m_{1}^{2})N^{2}}{4E_{0}^{2}},
\end{equation}%
\begin{equation}
P_{1}\approx P_{0}(1-\frac{m_{0}^{2}-m_{1}^{2}}{4E_{0}^{2}}N^{2})\text{,}
\end{equation}%
\begin{equation}
L_{2}\approx L_{0}\frac{(m_{0}^{2}-m_{1}^{2})N^{2}}{4E_{0}^{2}},
\end{equation}%
\begin{equation}
L_{1}\approx L_{0}-\frac{(m_{0}^{2}-m_{1}^{2})N^{2}}{4E_{0}^{2}}L_{0}.
\end{equation}

It is particle 2 that escapes but it has almost vanishing energy: the closer
point of decay to the horizon, the smaller the energy.

\subsubsection{Scenario II}

Both particles fall in a black hole. Then, it follows from (\ref{L1II}) - (%
\ref{P1II}) that near the horizon%
\begin{equation}
L_{1}\approx \frac{L_{0}}{2m_{0}^{2}}(b_{1}-\sqrt{d}-\sqrt{d}\frac{%
N^{2}m_{0}^{2}}{2E_{0}^{2}})\text{,}
\end{equation}%
\begin{equation}
L_{2}\approx \frac{L_{0}}{2m_{0}^{2}}(b_{2}+\sqrt{d}+\sqrt{d}\frac{%
N^{2}m_{0}^{2}}{2E_{0}^{2}})\text{,}
\end{equation}%
\begin{equation}
P_{1}\approx \frac{P_{0}}{2m_{0}^{2}}(b_{2}+\sqrt{d}+\frac{\sqrt{d}N^{2}}{%
2E_{0}^{2}}\text{,}
\end{equation}%
\begin{equation}
P_{2}\approx \frac{P_{0}}{2m_{0}^{2}}(b_{2}-\sqrt{d}-\frac{N^{2}m_{0}^{2}%
\sqrt{d}}{2E_{0}^{2}})\text{,}  \label{p1n}
\end{equation}%
\begin{equation}
E_{2}\approx \frac{E_{0}}{2m_{0}^{2}}(b_{2}-\sqrt{d}+\sqrt{d}N^{2}\frac{%
m_{0}^{2}}{2E_{0}^{2}})\text{,}
\end{equation}%
\begin{equation}
E_{1}\approx \frac{E_{0}}{2m_{0}^{2}}(b_{1}+\sqrt{d}-\sqrt{d}N^{2}\frac{%
m_{0}^{2}}{2E_{0}^{2}}).
\end{equation}

Now, for $m_{2}\neq 0$ both particles have comparable energies.

It is worth noting that, according to (\ref{c1}), in scenario I only
massless particle can escape that makes this scenario possible as exception.
This can be explained as follows on the basis of eq. (\ref{v20}). In the
center of mass frame (that coincides with particle 0) particles should move
after decay in opposite directions. To return to the laboratory frame, one
has to make a local Lorentz boost. For a massive particle, near the horizon
this boost grows indefinitely and overcomes an initial velocity, so particle
is drifted towards a black hole. However, if a particle is massless, the
velocity is equal to the speed of light from the very beginning and the
boost with large but finite value cannot change the direction of propagation
(if we compare both frames).

And, in scenario II, the situation is opposite: according to (\ref{c2}),
only in the massless case particle 2 cannot fall in a black hole. The reason
is the same. Particle 1 in scenario II falls in a black hole, so in the
center of mass frame particle 2 moves in the outward direction. If particle
2 is massless, even big but finite boost cannot overcome this, and particle
2 cannot move inwardly, scenario II fails. Instead, we can take just
particle 1 to be massless. Then, nothing wrong occurs and scenario II is
satisfied.

An important reservation is in order. These reasonings are valid for static
black holes but fail in the ergosphere of a rotating black holes. There,
another factors come into play and make it impossible for particle 2 to move
outwardly - see above discussion before eq. (\ref{en1}).

It is worth stressing the difference between rotating and static black holes
in the context under discussion. For rotating black holes, scenario I is
forbidden and particle 2 can escape only after preliminary bounce from the
potential barrier. For static black holes, this scenario is possible even if
decay occurs on the horizon. But, in the latter case, one should take the
sign "minus" before a square root in the efficiency (\ref{efst}). Then, in
the near-horizon limit, $\eta \rightarrow 0$. From another hand, scenario TP
in the near-horizon limit is impossible for static black holes and neutral
particles since it is inconsistent with the condition $E_{0}\geq m_{0}$.

\subsection{Comparison with examples from literature}

The formalism developed above enables us to reproduce the results for
particular metrics.

\subsubsection{Schwarzschild metric}

In this case, our eqs. (\ref{L2stat}), (\ref{L1stat}) agree with eq. (110)
of \cite{rocket}. It follows from (\ref{e2st}) that for scenario I%
\begin{equation}
\eta _{Sch}=\frac{E_{2}}{E_{0}}=\frac{b_{2}}{2m_{0}^{2}}-\frac{\sqrt{d}\sqrt{%
1-\frac{N^{2}}{\varepsilon _{0}^{2}}}}{2m_{0}^{2}}\text{,}
\end{equation}%
where $\varepsilon _{0}=\frac{E_{0}}{m_{0}}$.

There is serious disagreement between \ our formulas and \cite{ju}. In that
paper, the authors considered the efficiency according to their eqs. (34),
(35). If one neglects there the electric charges, it follows from the
aforementioned equations that $E_{3}\approx E_{1}$ for decay that occurs on
the horizon, so the efficiency $\eta \approx 1$. But, according to our
treatment above, this consideration applies only to a particle that falls in
a black hole. Meanwhile, the efficiency of the process should be evaluated
for a particle that escapes. Such a particle has almost vanishing energy (%
\ref{e2n}) of the order $N^{2}$, so the efficiency $\eta \ll 1$.

\subsubsection{Magnetic field}

The Penrose process in the electromagnetic field is a separate issue beyond
the scope of the present article. We only discuss very briefly for
illustration purposes motion of neutral particles in the Ernst background 
\cite{ernst}, 
\begin{equation}
N^{2}=(1-\frac{2M}{r})(1+\frac{B^{2}r^{2}}{4}\sin ^{2}\theta )\text{.}
\end{equation}%
Full-fledged picture of particle motion in this space-time is quite \
nontrivial and includes chaotic behavior \cite{kar92}. However, for our
illustrative purposes it is quite sufficient to consider the case of motion
within the plane $\theta =\frac{\pi }{2}$, according to the subject of the
present article. Fore mor details about properties of particle motion
(charged or neutral) in this background see \cite{78}.

Then, for motion in the plane $\theta =\frac{\pi }{2}$, in scenario I
according to (\ref{e21n}) we have 
\begin{equation}
E_{2}=\left( \frac{b_{2}}{2m_{0}^{2}}\right) (E_{0}-\sqrt{%
E_{0}^{2}-m_{0}^{2}N^{2}}),
\end{equation}%
\begin{equation}
\eta =\frac{E_{2}}{E_{0}}=\frac{b_{2}}{2m_{0}^{2}}-\frac{\sqrt{d}}{%
2m_{0}^{2}\varepsilon }\sqrt{\varepsilon ^{2}-(1-\frac{2M}{r_{0}})(1+\frac{%
B^{2}r_{0}^{2}}{4})\text{.}}
\end{equation}

Here, it is assumed that $\varepsilon =\frac{E_{0}}{m_{0}}>N(r_{0})$. It is
seen that for given $r_{0}$, $\eta >\eta _{Sch}$ where $\eta _{Sch}\equiv
\eta (B=0)$ corresponds to the Schwarzschild metric.

If $B\rightarrow 0$,%
\begin{equation}
\eta =\eta _{Sch}+\frac{\sqrt{d}}{2m_{0}^{2}\varepsilon }\frac{(1-\frac{2M}{%
r_{0}})\frac{B^{2}r_{0}^{2}}{2}}{\sqrt{\varepsilon ^{2}-(1-\frac{2M}{r_{0}})}%
}+O(B^{4}).
\end{equation}

It is worth noting that the Penrose process in this background was
considered quite recently in \cite{sh}. In our view, the corresponding
result for efficiency described by Eq. (31) in the aforementioned paper is
incorrect. It gives $\eta =0$ in the neutral case $q=0$, $Q=0$, $B=0$
instead of the Schwarzschild value $\eta _{Sch}$.

\section{ Alternative approach\label{two}}

In this section, we compare two seemingly different approaches for the
description of particle decay. In this section, we proceed along the lines
of Refs. \cite{dad1}, \cite{win}, \cite{j}, \cite{ju}, \cite{sh}, \cite{nz}
and demonstrate that they are equivalent to general formulas derived above.

One can write

\begin{equation}
p^{t}=-\frac{E}{g_{tt}+\Omega g_{t\phi }}\text{,}
\end{equation}%
\begin{equation}
p^{\phi }=p^{t}\Omega .
\end{equation}%
It is easy to check using (\ref{ome}) that 
\begin{equation}
-p_{t}=E,\text{ }
\end{equation}%
as it should be.

For a circle orbit, the equation $g_{\mu \nu }p^{\mu }p^{\nu }=-m^{2}$ gives
us%
\begin{equation}
\Omega =\frac{-g_{t\phi }(\varepsilon ^{2}+g_{tt})+\sqrt{(\varepsilon
^{2}+g_{tt})\varepsilon ^{2}g_{\phi }N}}{(g_{t\phi }^{2}+g_{\phi
}\varepsilon ^{2})}.
\end{equation}%
As on the circle orbit $\dot{r}=0\,$, this equation should coincide with (%
\ref{o0}) where $\Omega $ in the turning point is given. After some
algebraic manipulation one can confirm that this is indeed the case.

Using the conservation laws, we can write

\begin{equation}
p_{(0)}^{t}Y_{(0)}=p_{(1)}^{t}Y_{(1)}+p_{(2)}^{t}Y_{(2)}\text{,}
\end{equation}%
\begin{equation}
p_{(0)}^{t}\Omega _{(0)}=p_{(1)}^{t}\Omega _{(1)}+p_{(2)}^{t}\Omega _{(2)}%
\text{,}
\end{equation}%
where 
\begin{equation}
Y\equiv g_{00}+\Omega g_{t\phi },  \label{Y}
\end{equation}%
one obtains for the efficiency%
\begin{equation}
\chi =\frac{E_{2}}{E_{0}}=\frac{P_{(2)}^{t}Y_{(0)}}{Y_{(2)}P_{(0)}^{t}}
\label{e22}
\end{equation}%
that is equivalent to 
\begin{equation}
\chi =\frac{\Omega _{0}-\Omega _{1}}{\Omega _{2}-\Omega _{1}}\frac{Y_{0}}{%
Y_{2}}
\end{equation}%
in agreement with Refs. mentioned in the beginning of this Section.

One can check that (\ref{ef}) agrees completely with (\ref{e22}).

Thus both approaches give the same results. The approach based on formulas (%
\ref{L2I}) - (\ref{P1II}), has the advantages in that (i) it operates
directly with characteristics which are given in the problem as initial
conditions (energy $E_{0}$ and angular momentum $L_{0}$), (ii) it gives not
only energies of particles 1 and 2 but also their angular momenta, (iii) is
applicable not only in the point of decay or on a circle orbit.

\section{Summary and conclusions\label{sum}}

We considered the background of a rotating axially symmetric black hole.
Classification of scenarios of decay to two particles is suggested. We
mainly restricted ourselves by the case when both particles are ejected in
the parallel direction with respect to the motion of a parent particle. It
is this particular case which is of the maximum interest since it enables us
to obtain maximum efficiency. In particular, we showed that inside the
ergoregion there is severe restriction on some scenarios of this type. Thus,
it is shown that if decay occurs in a generic point (not a turning point for
radial motion) inside the ergoregion, both new particles 1 and 2 move after
decay in the same radial direction that coincides with direction of a parent
particle 0.

We argued that description of the Penrose process can be taken directly from
the formulas derived by Wald many years ago. However, these formulas were
written in terms of the velocities of particles with respect to a parent
particle 0. Meanwhile, we transformed the corresponding formulas, giving
them in terms of particle masses directly. In particular, in this manner, we
found the expressions for the gamma factor of relative motion of particles
before and after decay. Scenarios \ with decay in the turning point are also
considered in a model-independent way. We also discussed one more approach
for description of decay, popular in recent years, and showed its
equivalence to ours (with advantage that our approach gives the answer
explicitly in terms of characteristics of particle 0 and all three particle
masses).

Our results enabled us to reject some misleading formulas in previously
published works where physically incorrect results appeared due to confusion
between particles that fall in a black hole and those that escape.
Especially, this concerns the decay near the horizon, where it is important
to know the relation between the energy of escaping particle and direction
of its motion.

We also considered in detail possible variants of motion in the angular
direction. In doing so, the angular momenta of particles 1 and 2 are found
explicitly. We traced in detail how the "plus" and "minus" signs are
interwoven in different scenarios.

We collected and enlarged the results, how efficiency depends on the
scenario, location of a point where decay occurred, and particle passes.
Some part of corresponding statements were already known but existed as some
folklore scattered over literature and discussed for different models. Now,
they are given more accurate substantiation, also in a model-independent way.

Of interest is to apply the developed approach to decay processes in the
external electromagnetic field and consider astrophysical applications.

\section{Acknowledgment}

I am grateful to Yuri Pavlov for interest to this work and useful
discussion. I also thank Naresh Dadhich for useful correspondence and some
references. I thank for hospitality Institute of Theoretical Physics of
Charles University (Prague) and CENTRA (Lisbon). I acknowledge financial
support from grant No. GA\v{C}R 21/11268S of the Czech Science Foundation
and Funda\~{c}ao para a Ci\^{e}ncia e a Tecnologia - FCT, Portugal, project
No. UIDB/00099/2020.

\end{document}